\documentclass[prl,twocolumn,amsmath,amssymb,showpacs]{revtex4}
\usepackage{wasysym}
\usepackage{epsf}
\usepackage{graphicx}
\usepackage{bm}
\begin{document}
\title{Asymmetric polarity reversals, bimodal field distribution,
and coherence resonance in a
spherically symmetric mean-field dynamo model}
\author{Frank Stefani and Gunter Gerbeth}
\affiliation{Forschungszentrum Rossendorf\\
P.O. Box 510119, D-01314 Dresden, Germany}


\begin{abstract}
Using a mean-field dynamo model with a spherically symmetric
helical turbulence parameter $\alpha$ which is dynamically quenched
and disturbed by additional noise, the basic features of
geomagnetic polarity reversals are shown to be generic
consequences of the dynamo action in the vicinity of exceptional points
of the spectrum.
This simple paradigmatic model yields long periods of constant polarity
which are interrupted by self-accelerating field decays leading to
asymmetric polarity reversals. It shows the
recently discovered
bimodal field distribution, and it gives a natural explanation of the
correlation between polarity persistence time and field strength.
In addition, we find typical features of
coherence resonance in the dependence of the persistence
time on the
noise.
\end{abstract}


\maketitle

The Earth's magnetic field is known to undergo
irregular polarity reversals, with a mean reversal rate
that
varies from zero in the Permian and
Cretaceous supercrons
to (4-5) per Myr in the present \cite{MERR}.
Typically, these reversals have an
asymmetric (saw-toothed)
shape, i.e.
the field of one polarity decays slowly and recovers very
rapidly with the
opposite polarity, possibly to rather high intensities
\cite{VALET-MEYN-BOGU}.
A general correlation between the persistence time and the
field intensity has also been suspected since long
\cite{COX-TARDUNO}.
A recent observation concerns the bimodal distribution
of the Earth dipole moment with two peaks at about
4 $\times$ 10$^{22}$ Am$^2$
and at about twice that value \cite{HELLER}.

The explanation of these
phenomena represents a great challenge for dynamo
theory and numerics. Remarkably,
the last decade has seen three-dimensional numerical
simulations of
the geodynamo with sudden
polarity reversals as one of the most impressive results
(cf. \cite{GLRO-ANDCO} for a recent overview).

Despite the fact that those simulations exhibit many
features of the Earth's magnetic field quite well, and
{\it because they do so}
in parameter regions
(in particular for the Ekman and the magnetic Prandtl number)
that are
far away from the real ones, there is a complimentary tradition to
identify the
essential ingredients of reversals within the framework of
simplified dynamo
models. This has
been done, e.g., in the tradition
of the celebrated Rikitake dynamo model of two coupled
disk dynamos \cite{RIKI-FRANCK-HIDE}.
Another approach has been pursued by Hoyng and collaborators
\cite{HOYNG} who studied a prototype
nonlinear mean-field dynamo model which is reduced
to an equation system for
the amplitudes of the non-periodic axisymmetric dipole mode
and for one periodic overtone under
the influence of stochastic forcing.
Interestingly, even this simple model shows
sudden reversals and a
Poissonian distribution of the polarity persistence time.
However, an essential ingredient of this model to account for
the correct reversal duration and persistence time is
the use
of a large turbulent resistivity which is hardly justified, at
least not by the recent dynamo experiments \cite{RMP}.

\begin{figure}
\begin{center}
\epsfxsize=8.0cm\epsfbox{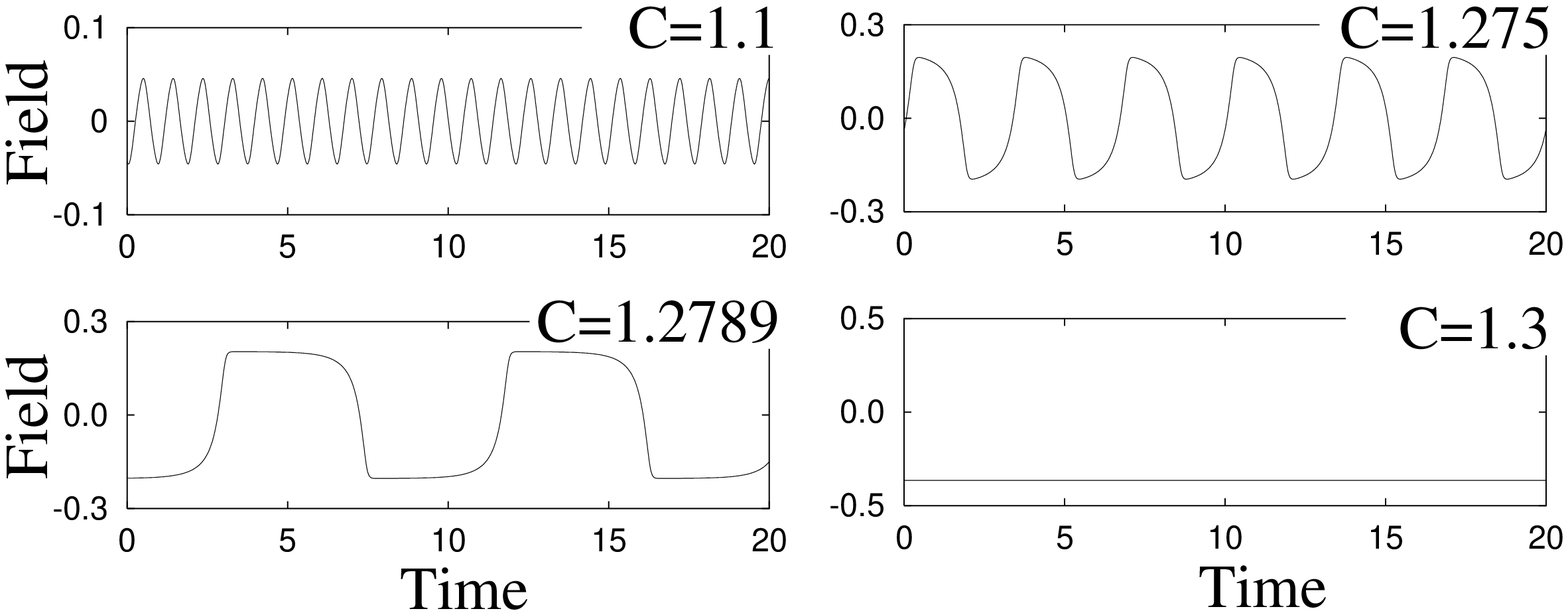}
\end{center}
\caption{Magnetic field evolution for vanishing noise ($D=0$)
and different values of $C$. ''Field'' corresponds to $s_1(r=0.95)$.}
\end{figure}

Sarson and Jones \cite{SAJO} had noticed the importance of the
transition from non-oscillatory to oscillatory states for reversals
to occur. It is our goal to understand
this process in more detail by studying a simple mean-field dynamo
model.
We focus on
the magnetic field dynamics
in the vicinity of ''exceptional points'' \cite{KATO-HEIS} of
the spectrum
of a non-selfadjoint operator.
We will show that the main characteristics of
Earth magnetic field
reversals can be attributed to the square-root character of the
spectrum in the vicinity of such exceptional points, where
two non-oscillatory eigenmodes coalesce and continue as an
oscillatory eigenmode.

Our starting point is the well known induction equation
${\dot{ \bm{B}}} ={\bm \nabla}
\times (\alpha {\bm{B}}) +
(\mu_0 \sigma)^{-1} \Delta {\bm{B}}$
for a  mean-field dynamo with a
helical turbulence parameter
$\alpha$  \cite{KRRA}, acting in a fluid with
electrical conductivity $\sigma$ within a sphere
of radius $R$. The magnetic field has to be
divergence-free,${\bm{\nabla}} \cdot {\bm{B}}=0$.
Henceforth,  we will
measure the length in units of $R$,
the time in units of $\mu_0 \sigma R^2$,
and
the parameter $\alpha$ in units of
$(\mu_0 \sigma R)^{-1}$.
Note that for the Earth  we get a
time scale $\mu_0 \sigma R^2 \sim 200$ Kyr,
giving a free decay time of 20 Kyr for
the dipole field.

\begin{figure}
\begin{center}
\epsfxsize=8.0cm\epsfbox{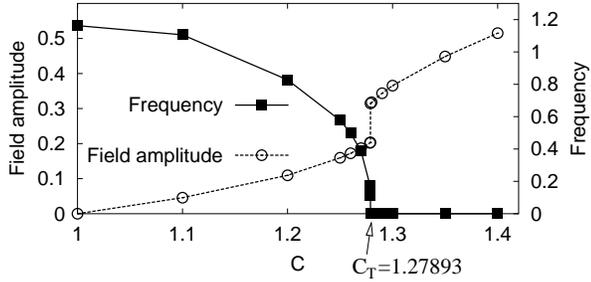}
\end{center}
\caption{Magnetic field amplitude and frequency for $D=0$
in dependence on $C$. Note the phase transition at $C_T=1.27893$.}
\end{figure}

We decompose ${\bm{B}}$ into a
poloidal and a toroidal part,
${\bm{B}}=-\nabla \times ({\bm{r}} \times
\nabla S)-{\bm{r}} \times
\nabla T $.
The defining scalars $S$ and $T$ are
expanded
in spherical harmonics of degree $l$ and order $m$
with the expansion coefficients
$s_{l,m}(r,t)$ and $t_{l,m}(r,t)$.
In order to
allow for very long simulations (to get
a good statistics for the persistence time and the
field amplitudes),
we consider an $\alpha^2$ dynamo with a
radially symmetric helical turbulence parameter $\alpha$.
In \cite{OSZI} we had shown that such simple $\alpha^2$ dynamos
can exhibit oscillatory behaviour in case that $\alpha(r)$
changes its sign along the radius, which is not unrealistic for the
Earth \cite{SOW-RUED}.
For spherically symmetric and isotropic $\alpha$,
the induction equation
decouples for each $l$ and $m$ into the following pair
of equations:
\begin{eqnarray}
\frac{\partial s_l}{\partial t}&=&
\frac{1}{r}\frac{d^2}{d r^2}(r s_l)-\frac{l(l+1)}{r^2} s_l
+\alpha(r,t) t_l \; ,\\
\frac{\partial t_l}{\partial t}&=&
\frac{1}{r}\frac{d}{dr}\left( \frac{d}{dr}(r t_l)-\alpha(r,t)
\frac{d}{dr}(r s_l) \right) \nonumber\\
 &&-\frac{l(l+1)}{r^2}
[t_l-\alpha(r,t)
s_l] \; .
\end{eqnarray}
These equations are independent of the order $m$, hence we
have skipped it in the index of $s$ and $t$.
The boundary conditions are
$\partial s_l/\partial r |_{r=1}+{(l+1)} s_l(1)=t_l(1)=0$.

In the following we restrict ourselves to the dipole field with $l=1$
(the influence of higher multipole fields will be considered
elsewhere). At first, we choose a particular radial profile $\alpha(r)$
which can yield an oscillatory behaviour \cite{OSZI}.
Magnetic field saturation is ensured by quenching the
parameter $\alpha(r,t)$ with the angular averaged dipole field
energy which
can be expressed in terms of $s_{1}(r,t)$ and $t_1(r,t)$. In addition
to that, we perturb the $\alpha$-profile by
noisy "blobs" which are assumed constant within a correlation
time $\tau$. In summary, $\alpha(r,t)$ takes the form:
\begin{eqnarray}
\alpha(r,t)&=&C \frac{-21.5+426.4 \; r^2-806.7 \; r^3+392.3 \; r^4}{1+
E \left[ \frac{2 s_1^2(r,t)}{r^2}+
\frac{1}{r^2}\left( \frac{\partial (r s_1(r,t))}
{\partial r} \right)^2
+t_1^2(r,t) \right] }     \nonumber\\
&& +\xi_1(t) +\xi_2(t) \; r^2 +\xi_3(t) \; r^3+\xi_4(t) \; r^4 \; ,
\end{eqnarray}
where the noise correlation is given by
$\langle \xi_i(t) \xi_j(t+t_1)
\rangle = D^2 (1-|t_1|/\tau) \Theta(1-|t_1|/\tau) \delta_{ij}$.
$C$ is a normalized dynamo number measuring the overcriticality,
$D$ is the noise
amplitude, and $E$ is a constant measuring the inverse equipartition
field energy.

The equation system (1)-(3) is time-stepped
using an Adams-Bashforth method with a radial grid spacing
of 0.02 and a time step length of 2 $\times$ 10$^{-5}$.
For the following examples,
the correlation time $\tau$ has been set to 0.02,
and $E$ has been
chosen to be 0.01.

\begin{figure}
\begin{center}
\epsfxsize=8.0cm\epsfbox{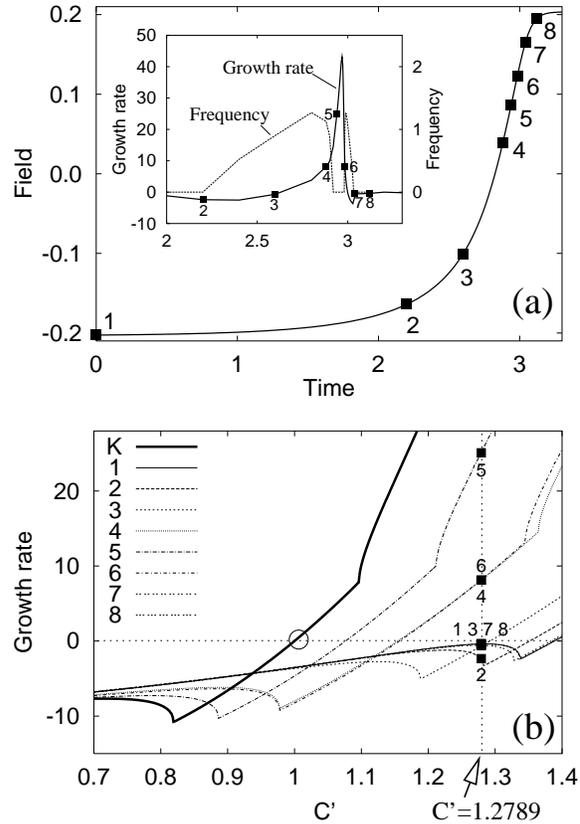}
\end{center}
\caption{(a) Details of the field evolution for $C=1.2789$ (cf. Fig. 1), with
growth rates and frequencies (inset) resulting from the
instantaneous, quenched $\alpha$ profiles.
(b) Growth rate curves for
instantaneous $\alpha$ profiles scaled by $C'$. ''K'' denotes the
kinematic (unquenched) $\alpha$ profile with the (encircled) critical
point at $C'=1$.}
\end{figure}

For the noise-free case ($D=0$), we show in Fig. 1
the evolution of the magnetic field for different values
of $C$ (hereafter, ''field'' stands always for the
value of  $s_1$ at $r=0.95$). At the slightly overcritical
value  $C=1.1$ we observe a
nearly harmonic
oscillation with a frequency close to one. With increasing
$C$, this frequency decreases, and at the same time the
signal becomes
saw-toothed (or better: ''shark-fin shaped'')
and eventually rectangular. At
the critical point $C_T=1.27893$ a phase
transition to a
non-oscillatory dynamo occurs, which can also be identified in the
frequencies and field amplitudes shown in Fig. 2.

\begin{figure}
\begin{center}
\epsfxsize=8.0cm\epsfbox{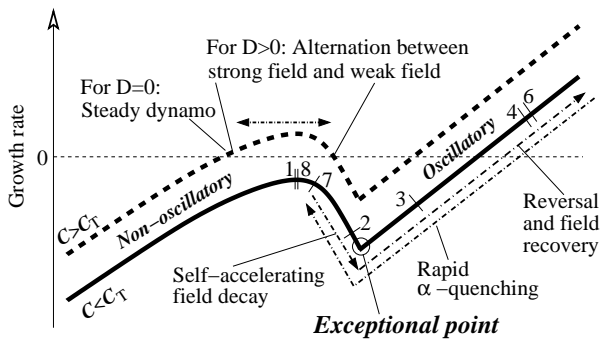}
\end{center}
\caption{Illustration of the various phases of reversals in the vicinity
of an exceptional point of the spectrum. In the noise free case
(D=0) oscillatory and
steady dynamos can be sharply distinguished.
For $D>0$, we get a bimodal behaviour around the maximum of the
non-oscillatory branch.}
\end{figure}

In Fig. 3, we analyze the field evolution for $C=1.2789$
in detail. Figure 3a shows the field during one reversal,
together with the growth rate and the frequency
that result from the {\it instantaneous}, quenched $\alpha$-profile.
Eight of these profiles (at the moments 1-8), together with
 the unquenched profile,
are scaled by a factor $C'$ yielding the growth rate curves of
Fig. 3b.
These curves may help us to
identify the following phases for a reversal:
a slowly starting, but self-accelerating field decay in the
non-oscillatory branch (points 1 and 2), followed
by the actual polarity reversal in the oscillatory branch
(between points 3 and 4), a fast
increase of the field (points 4-6) which results in a
rapid $\alpha$-quenching. This rapid quenching drives the system back
to the point 8 which basically
corresponds to the initial point 1, only with the opposite field
polarity.
A peculiarity of our particular model is the existence
of a second exceptional point
beyond which the dynamo becomes non-oscillatory again (between points 5 and 6),
which is, however, not crucial for the indicated reversal process.

In an attempt to simplify this reversal picture, we
consider in Fig. 4 all growth rate curves of
Fig. 3b as being
collapsed into a single one (for that purpose we
neglect their slight shape differences and consider only
a shift).
By this ''optical trick'', the various phases of reversals
can be visualized as a ''move'' of the actual
growth rate point relative to this collapsed
growth rate curve.
Figure 4 explains also the
phase transition between oscillatory and steady dynamos
and the discontinuity of the field
amplitude from Fig. 2.
The local maximum of the non-oscillatory branch
of the growth rate curve
is unstable and
repels the dynamo in two different
directions, depending on whether the growth rate there is
negative or positive.
If it is negative, the field decays, we get weaker $\alpha$ quenching,
and a subsequent reversal. If it is positive, the field increases,
we get stronger
$\alpha$ quenching, hence the system moves to the left of the maximum.

\begin{figure}
\begin{center}
\epsfxsize=8.0cm\epsfbox{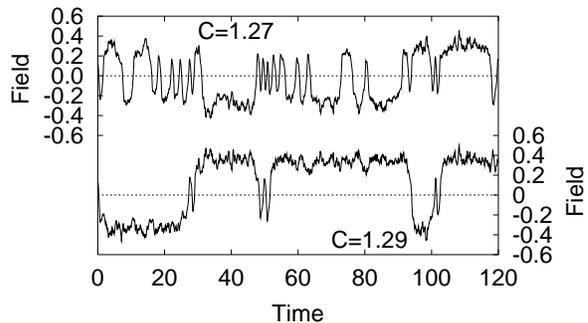}
\end{center}
\caption{Time series for $C=1.27$ and $C=1.29$ at $D=0.5$.}
\end{figure}

What happens now if we switch on the noise? The influence of the
noise is quite different for values of $C$ below and above $C_T$.
For $C>C_T$, the noise is the
only possibility
to trigger a move from left of the local maximum to the right,
hence the persistence time will decrease from infinity
to some finite value. For $C<C_T$, the noise will
sometimes push the maximum above the zero growth rate line allowing
the system to jump to the left of the maximum where it
can stay for a while. Hence, we will get an increase of
the persistence time.

For a moderate noise intensity $D=0.5$,
we depict in Fig. 5 the time evolution for the values
$C=1.27$ and $C=1.29$ which are slightly below and above $C_T$,
respectively.
First of all, a drastic difference between the persistence times
is still visible.
The duration of a reversal, however, is quite identical for both
values of $C$, although its sensible definition is not obvious.
With a side view on the real geomagnetic field (for which the
dipole component  should decay
approximately to the strength of the non-dipole components
before a reversal can  be identified), we tentatively
define the reversal
duration as the time
the field spends between $\pm 1/4$ or $\pm 1/10$ of its mean
intensity. In the first case we get a reversal duration of 30
Kyr, in the
second case of 12 Kyr, which comes close to the observational facts.

\begin{figure}
\begin{center}
\epsfxsize=8.0cm\epsfbox{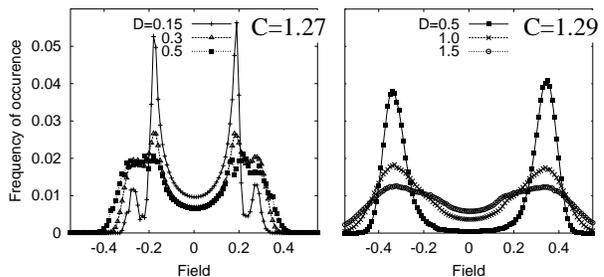}
\end{center}
\caption{Field histogram for $C=1.27$ and $C=1.29$.}
\end{figure}

For both values, $C=1.27$ and $C=1.29$, a bimodal behaviour
of
the  dynamo can be observed in the
field histograms  of Fig. 6. For $C=1.27$ and $D=0.15$, we observe a
clear double peak on both polarity sides, centered approximately at
$\pm 0.18$ and $\pm 0.27$.
Evidently, the dynamo is mostly in the weak field state
(right of the maximum, cp. Fig. 4),
with some ''excursions'' to the strong field state (left of
the maximum).
With increasing $D$ this double peak is smeared out,
leaving a very broad maximum for $D=0.5$.
The three histograms for $C=1.29$ show a  maximum at
the strong field value and (although not a maximum)
a pronounced flattening at the weak field value.
This means, the dynamo is mostly in
the strong field state, with some intermediate stopovers in the
weak field state, from where it can start a reversal.

\begin{figure}
\begin{center}
\epsfxsize=8.0cm\epsfbox{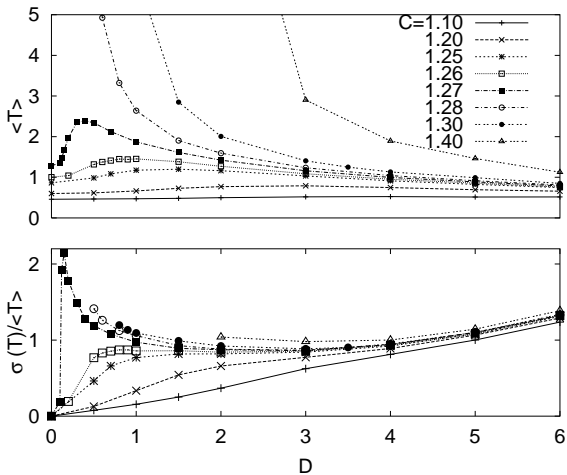}
\end{center}
\caption{Persistence time  $\langle T \rangle$  and its normalized
standard deviation
$\sigma(T)/\langle T \rangle$
for different $C$
in dependence on $D$. For each point, a minimum of
10$^4$ reversals was used.}
\end{figure}

An overview about the mean persistence time $\langle T \rangle$ and
its normalized standard deviation $\sigma(T)/\langle T \rangle$
in dependence on $C$ and $D$ is given in Fig. 7.
Note the drastic difference that small values of noise have on
the dynamo behaviour for $C<C_T$ and $C>C_T$.
The normalized standard deviation, at least for
the curves with $C>1.26$,  has a clear minimum
around $D=3$ which represents a typical feature of
{\it coherence} resonance \cite{KURTHS}. We only note
here that recently
{\it stochastic} resonance models have been discussed with view
on a
possible triggering of field reversals by the 100 Kyr period
of the Earth orbit eccentricity \cite{STOCH}.

To summarize, our simple spherically symmetric $\alpha^2$ dynamo model
shows that asymmetric polarity
reversals
and bimodal field distributions are generic features
of the dynamo behaviour
in the vicinity
of exceptional points. Using only the molecular resistivity of
the outer core, we get typical persistence times of the order of 200 Kyr
(and larger), and a typical reversal duration of 10-30 Kyr.

We point out that this model is not a physical model of the Earth dynamo,
in particular owing to the missing North-South asymmetry of $\alpha$.
Nevertheless, we
are convinced that the generic features of dynamos in the vicinity of
exceptional points as they were illustrated in this paper can
also be identified in much more elaborated dynamo models.

\section*{ACKNOWLEDGMENTS}

We thank U. G\"unther for fruitful discussions,
and J. Kurths for drawing our attention to the topic
of coherence resonance. This work was supported
by Deutsche Forschungsgemeinschaft in frame of  SFB 609
and grant No. GE 682/12-2.

\end{document}